\documentclass[sigconf]{acmart}

\usepackage{booktabs}
\usepackage{multirow}
\usepackage{enumitem}
\usepackage{graphicx}
\usepackage{array}
\usepackage{float}
\usepackage{stfloats}
\usepackage[table]{xcolor}
\usepackage{subcaption}
\AtBeginDocument{%
  }

\copyrightyear{2025}
\acmYear{2025}

\begin{document}

\title{HF-SID: High-Fidelity Semantic IDs for Generative Retrieval in Location-Based Services}
\author{Haowen Lin}
\authornote{These authors contributed equally to this research.}
\email{linhaowen.lhw@taobao.com}
\affiliation{%
  \institution{AMAP, Alibaba Group}
  \city{Beijing}
  \country{China}
}

\author{Jing Li}
\authornotemark[1]
\email{lijingsgy@mail.ustc.edu.cn}
\affiliation{%
  \institution{University of Science and Technology of China}
  \city{Hefei}
  \country{China}
}

\author{Zhibin Hao}
\authornotemark[1]
\email{hzb24@mails.tsinghua.edu.cn}
\affiliation{%
  \institution{Tsinghua University}
  \city{Beijing}
  \country{China}
}

\author{Fangye Wang}
\email{wangfangye.wfy@alibaba-inc.com}
\affiliation{%
  \institution{AMAP, Alibaba Group}
  \city{Beijing}
  \country{China}
}

\author{Lihui Su}
\email{slh@pku.edu.cn}
\affiliation{%
  \institution{AMAP, Alibaba Group}
  \city{Beijing}
  \country{China}
}

\author{Song Yang}
\email{song.yangs@alibaba-inc.com}
\affiliation{
\institution{AMAP, Alibaba Group}
  \city{Beijing}
  \country{China}
}

\author{Xiaojiang Zhou}
\authornote{Corresponding author.}
\email{zhouxiaojiang.zxj@taobao.com}
\affiliation{%
  \institution{AMAP, Alibaba Group}
  \city{Beijing}
  \country{China}
}

\author{Pengjie Wang}
\email{pengjie.wpj@alibaba-inc.com}
\affiliation{%
  \institution{AMAP, Alibaba Group}
  \city{Beijing}
  \country{China}
}

\renewcommand{\shortauthors}{Lin et al.}
\begin{abstract}
Generative retrieval has attracted increasing attention in Location-Based Services (LBS), where each Point-of-Interest (POI) is represented as a Semantic ID (SID). As the SID is the only channel through which POI information reaches the generative model, whatever it fails to preserve is irrecoverable at decoding time, and LBS retrieval is especially sensitive to the fine-grained differences that existing SIDs blur. Specifically, (1) LLMs embed continuous coordinates discontinuously, so their numeric differences do not reflect true geographic distance; (2) dynamic numerical attributes differ vastly in scale, so an identical gap may be decisive for one attribute yet negligible for another; and (3) short text cannot convey hierarchical affiliation, as text-similar POIs may belong to different hierarchies. We therefore propose HF-SID, which restores geographic, numerical, and structural fidelity at the representation stage, before any information is committed to a discrete code. It transforms coordinates into a continuous 3D Cartesian form and encodes each numerical value as a single unit, consolidated inside the LLM by Geo-CPT and Num-CPT with type-aware embeddings; a Structure-based Contrastive Learning objective, applied only to the last-layer residual, then separates co-located POIs that share a coarse tag but differ at the fine level. Because these mechanisms enrich the representation rather than lengthen the identifier, HF-SID uses a 3-token SID at no extra decoding cost. On a large-scale industrial dataset it reduces the average intra-SID geographic distance by 95.9\% and improves Hit@200 by 3.66 points over the strongest baseline, with online A/B gains of +6.74\% PV\_CVR and +6.03\% UV\_CVR. We further release \textbf{AMap-S}, a large-scale real-world POI dataset.
\end{abstract}


\maketitle

\section{Introduction}

Large language models (LLMs)~\cite{naveed2025comprehensive, minaee2024large} have achieved remarkable success across many tasks, driving the emergence of Generative Retrieval (GR)~\cite{li2025matching, pang2025generative, chen2025onesearch}. A central challenge in GR is transforming diverse item attributes into Semantic IDs (SIDs)~\cite{TIGER, deng2025onerec, zhou2025onerec}, since the SID is the only channel through which item information reaches the generative model: whatever it fails to preserve is permanently invisible at decoding time. This bottleneck is acute in Location-Based Services (LBS)~\cite{hiergr, GeoGR, lin2025spacetime}, where retrieval outcomes hinge on fine-grained POI attributes that mainstream LLMs cannot perceive precisely: they fragment coordinates into digit sub-tokens~\cite{liu2026geography}, are insensitive to the scale of numerical attributes~\cite{li2025exposing, ni2026numeracy}, and infer structure only from surface text~\cite{ROS}. The central question for LBS generative retrieval is therefore not how to \textit{organize} SIDs, but how to make them \textit{high-fidelity}. Existing POI-oriented SID methods pursue this goal but intervene too late. They append discrete spatial tokens such as Geohash or S2 Cell IDs as geographic prefixes~\cite{GENPOI, ROS}, inject geographic offsets during codebook quantization~\cite{ProGEO}, or fine-tune the embedding model with co-visit contrastive signals~\cite{GeoGR}. All of them operate on representations whose precision is already lost, since it vanishes the moment the tokenizer splits a coordinate into digit fragments; no downstream grid, rotation, or clustering can recover a value that was never continuous. Moreover, they address geography alone, leaving numerical attributes and structural information to plain text, and pay for their gains with longer identifiers that increase autoregressive decoding cost. Three challenges must therefore be resolved before the identifier is formed.

\begin{figure}[t]
\setlength{\abovecaptionskip}{0.2cm}
    \setlength{\belowcaptionskip}{-0.4cm}
    \centering
    \includegraphics[width=0.40\textwidth]{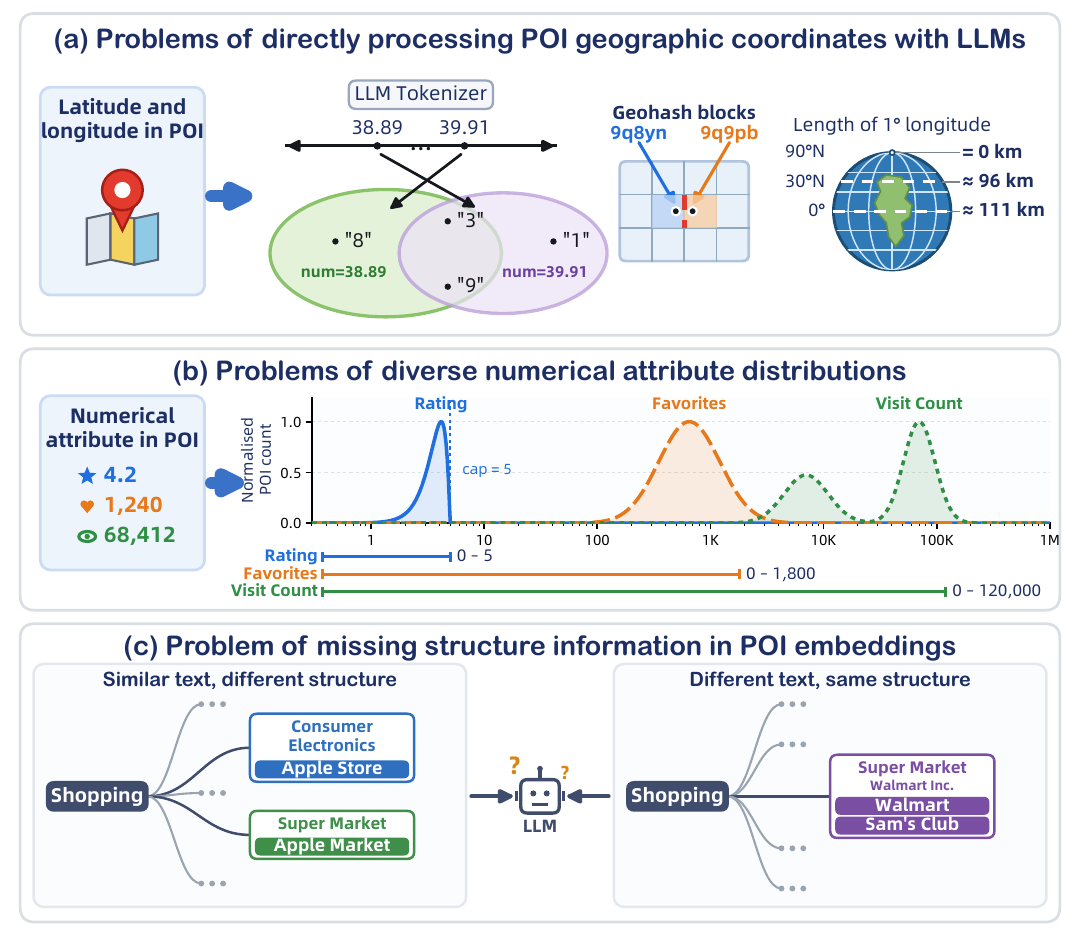}
    \caption{
        Three key challenges in encoding POI information into Semantic IDs: 
        geographic coordinates, dynamic numerical attributes, and hierarchical 
        structure.
    }
    \label{fig:challenges}
\end{figure}

\textbf{C1: Geographic coordinates.} In LBS retrieval a difference of merely $0.1^\circ$, roughly $10$\,km, already carries substantial business impact, yet LLMs mishandle coordinates in three ways (Figure~\ref{fig:challenges}(a)). Their numeric encoding is discontinuous~\cite{geonum, yang2025numbercookbooknumberunderstanding, li2025exposing}, so numerically close coordinates may land in distant embedding regions. Discrete grid encodings such as GeoHash~\cite{liu2014geohash} and S2~Cell~\cite{veach2017s2}, which some methods directly adopt as SID prefixes~\cite{GENPOI, ROS}, impose hard cell boundaries where two geographically adjacent POIs can be assigned entirely unrelated codes simply because a boundary falls between them. And coordinates are angular: the same longitude difference spans about $111$\,km at the equator but far less at high latitudes.

\textbf{C2: Dynamic numerical attributes.} POIs also carry ratings, average prices, favorite counts and so on, whose scales differ by orders of magnitude, so an identical gap means entirely different things (Figure~\ref{fig:challenges}(b)): $0.5$ in rating (range $0$--$5$) separates a well-regarded POI from a mediocre one, whereas $0.5$ in visit count (range $0$--$120$K) is indistinguishable from noise~\cite{guo2024embeddingcollapsescalingrecommendation}. Prior work~\cite{ProGEO, LGSID, GeoGR} serializes them as plain text without type awareness, forcing one shared numeric subspace to absorb all of them.

\textbf{C3 Hierarchical structure.} Each POI carries a system-annotated two-level tag such as ``Shopping/Consumer Electronics'', and sits inside a geographic unit such as a business district~\cite{gao2022geobert}, neither of which is recoverable from surface text. Tags may agree at the coarse level yet diverge at the fine level: ``Apple Store'' and ``Apple Market'' are textually almost identical and share the coarse tag ``Shopping'', yet belong to ``Consumer Electronics'' and ``Super Market'' (Figure~\ref{fig:challenges}(c)). A business district likewise hosts many heterogeneous shops that proximity cannot separate, and stronger geographic modeling only aggravates this: the more tightly co-located POIs are grouped, the less distinguishable they become.

To address these challenges, we propose HF-SID, which restores fidelity at the representation stage, before any information is committed to a discrete code. For C1, we encode each coordinate as a single continuous unit through a dedicated numerical encoder~\cite{geonum} and transform spherical coordinates into a 3D Cartesian form, which replaces hard grid boundaries with a continuous space in which Euclidean distance is monotonic in true geographic distance; Geo-CPT then consolidates this geometry inside the LLM through pairwise-distance and nearest-pair reasoning samples. For C2, we give each attribute type an independent Type Embedding that projects it into its own subspace, and Num-CPT trains the model on cross-attribute comparison samples so that a magnitude is always read relative to the attribute it belongs to. For C3, Structure-based Contrastive Learning treats two POIs as positives only when both tag levels match exactly, making coarse-level-only matches hard negatives. We apply it solely to the residual before the last quantization layer, since the first two layers have by then collapsed co-located POIs together while the fidelity they established remains untouched. As these mechanisms enrich the representation rather than append tokens to the identifier, HF-SID achieves this with a 3-token SID, adding no decoding step over the plain quantization baseline. It cuts the average intra-SID geographic distance by $95.9\%$ and lifts Hit@200 by $3.66$ points on a large-scale industrial dataset, with consistent online gains after deployment.

In summary, our main contributions are as follows:
\begin{itemize}[leftmargin=1em]
    \item We identify that existing POI SID methods lose geographic, numerical, and structural fidelity \textit{before} quantization, and propose HF-SID, which restores all three at the representation level while keeping the SID length unchanged.

    \item We introduce a unified numerical representation that encodes coordinates and dynamic attributes as single continuous units, combining a 3D Cartesian transformation with a numerical encoder and per-type type embeddings, and align it with the LLM through two-stage CPT, namely Geo-CPT for spatial reasoning and Num-CPT for scale-aware attribute comparison.

    \item We design Structure-based Contrastive Learning, which separates POIs that are co-located or share only a coarse tag but differ at the fine-grained level, and apply it only to the residual of the last quantization layer so that this refinement does not disturb the geographic signal encoded at coarser levels.

    \item We open-source AMap-S, a large-scale real-world dataset of POIs and trajectories, and validate HF-SID with large-scale offline experiments and a production deployment on AMap.
\end{itemize}

\section{Related Work}

\textbf{Generative Retrieval} reformulates item retrieval as a sequence-to-sequence task in which a language model directly generates structured item identifiers~\cite{DSI, NCI}. Its central design choice is the Semantic ID (SID), a short discrete code that must be both generatable by the LLM and semantically faithful to the item. SIDs are commonly obtained by Residual-Quantized Variational AutoEncoders or hierarchical K-Means over item embeddings~\cite{TIGER, LC-Rec, rqkmeans}, yielding a coarse-to-fine codebook sequence whose prefixes are shared by similar items, which lets constrained beam search traverse a billion-scale item space and underpins industrial deployments~\cite{deng2025onerec, zhou2025onerec}. A direct consequence is that whatever an item embedding fails to encode is irreversibly discarded once quantization is applied, so embedding fidelity, rather than codebook design, is the decisive factor. POI scenarios break down precisely here: text-driven SIDs assign nearly identical codes to semantically similar but spatially distant POIs, while adjacent POIs of different categories share no prefix at all, causing geographical hallucination during generation.

\textbf{Attribute Enhanced Semantic ID.} To inject attribute awareness into POI-oriented SIDs, existing work follows three routes. \textit{Discrete spatial tokenization} prepends quantized location codes to the semantic code, using S2 Cell encodings as hierarchical geographic prefixes~\cite{ROS}, Geohash-based Geographic IDs under trie-constrained decoding~\cite{GENPOI}, or hierarchical geographic tokenization with residual quantization~\cite{LGSID}. \textit{Token-level continuous injection} instead perturbs embeddings, through the Geographic Position Embedding of GenPOI~\cite{GENPOI} and the geo-centroid Geo-RoPE frame of Pro-GEO~\cite{ProGEO}. \textit{Behavior-driven alignment} reshapes the embedding space with log-mined supervision, contrasting spatiotemporal co-visit pairs with EM-style~\cite{dempster1977maximum} SID refinement~\cite{GeoGR}, or adding a diversity loss to RQ-VAE~\cite{lee2022autoregressive} against semantic collapse~\cite{GNPR-SID}. All of them act after coordinates have already been serialized as text, and all treat geography as the only structured signal worth modeling. Numerical attributes fare no better: sub-word tokenization destroys numeric continuity and leaves LLMs unreliable at magnitude and comparison~\cite{yang2025numbercookbooknumberunderstanding, li2025exposing, ni2026numeracy}, and although dedicated encoders that embed a scalar as a single continuous unit largely restore this ability~\cite{geonum}, they have been evaluated on reasoning tasks rather than identifier construction. Closest to us, ReSID~\cite{liang2026rethinking} and HiD-VAE~\cite{fang2025hid} quantize structured recommendation fields and multi-granularity semantics, yet both target general recommendation and neither confronts the heterogeneous attribute scales and spatial hierarchy of POIs.

\section{Methods}
\begin{figure*}[h]
    \setlength{\abovecaptionskip}{0.1cm}
    \setlength{\belowcaptionskip}{-0.2cm}
    \centering
    \includegraphics[width=0.9\textwidth]{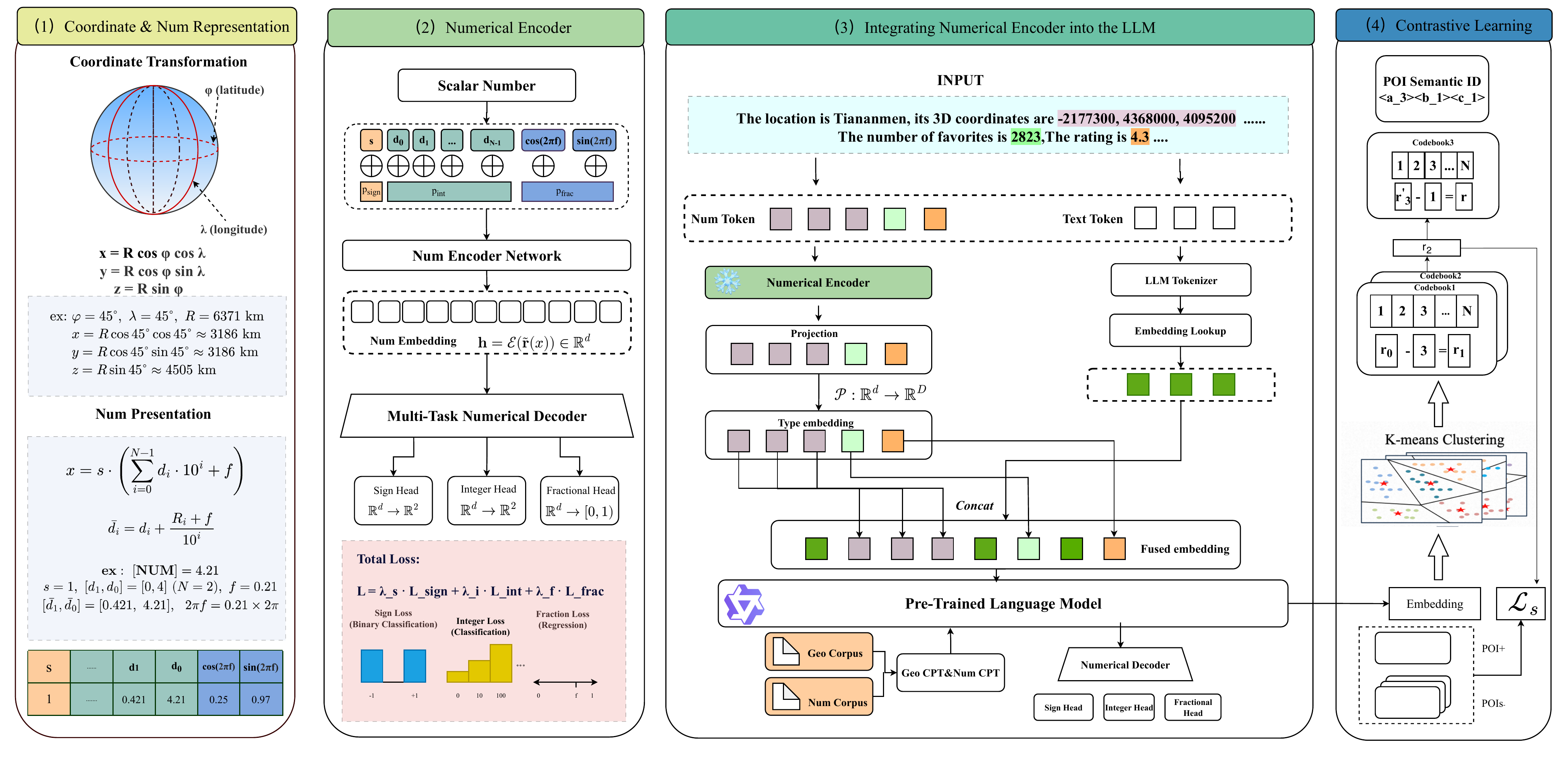}
    \caption{
        Overview of the HF-SID framework.
    }
    \label{fig:framework}
\end{figure*}

Given a POI $p$ with its textual description, HF-SID encodes it into a Semantic ID $\mathbf{SID}_p$, guided by a single principle: restore geographic, numerical, and structural fidelity at the representation level, before quantization commits the POI to a discrete code. As illustrated in Figure~\ref{fig:framework}, the framework consists of four stages. \textit{(1) Coordinate and Numerical Representation} (\S\ref{sec:geo_num}): coordinates are converted from spherical to 3D Cartesian form, and every numerical value in the description, coordinates included, is decomposed into a polar representation of its sign, per-digit magnitude, and an angular encoding of its fractional part, with these positions marked as numerical tokens. \textit{(2) Numerical Encoder} (\S\ref{sec:nce}): a dedicated encoder maps each polar representation to a continuous embedding, pretrained to reconstruct the sign, integer digits, and fractional part through three decoder heads, and then frozen and reused by all later stages. \textit{(3) Integrating the Numerical Encoder into the LLM} (\S\ref{sec:integrate}): each numerical embedding is projected into the LLM input space, offset by a Type Embedding that gives every attribute type its own subspace, and substituted at its original position to form a unified sequence together with text token embeddings; the LLM is then continually pre-trained by Geo-CPT and Num-CPT under a hybrid objective that applies cross-entropy at text positions and the numerical multi-task loss at numerical positions, and the hidden state of its last token becomes the POI representation. \textit{(4) HF-SID Generation with Structure-based Contrastive Learning} (\S\ref{sec:sid}): the POI representations are hierarchically quantized, where the first two layers cluster on geographic and numerical semantics and therefore give nearby POIs a shared prefix, while Structure-based Contrastive Learning refines the second-order residual before the third layer, so that POIs collapsed together by the coarser levels are separated by their tags in the final SID token.

\subsection{Coordinate and Numerical Representation}
\label{sec:geo_num}
POIs carry two categories of numerical information: geographic coordinates and dynamic numerical attributes. Standard LLM tokenizers fragment both into sub-tokens but destroy numerical continuity. And coordinates suffer from an additional problem, as raw latitude and longitude differences do not scale linearly with true distance. We therefore introduce two representations: a Cartesian transformation applied to the coordinates, and a unified polar decomposition applied to every numerical value, coordinates included.

\subsubsection{Coordinate Transformation}
We convert geographic coordinates from spherical to Cartesian form before they are embedded:
\begin{equation}
    x = R\cos\phi\cos\lambda, \quad
    y = R\cos\phi\sin\lambda, \quad
    z = R\sin\phi
    \label{eq:geo}
\end{equation}
where $\phi$ and $\lambda$ denote latitude and longitude, and $R$ is the mean Earth radius, so that $x, y, z \in \mathbb{R}$ are expressed in meters. Two properties follow. The representation is continuous across the antimeridian, since $\lambda = +179^\circ$ and $\lambda = -179^\circ$ now yield adjacent points. And the Euclidean distance between two POIs becomes $2R\sin\!\left(d/2R\right)$ for a great-circle distance $d$, a strictly monotonic and, at city scale, nearly linear function of true distance, which is far easier to model than raw latitude--longitude differences.

\subsubsection{Polar Representation of Numerical Values}

For every numerical value in the POI text, coordinates and dynamic attributes alike, we adopt a unified decomposition that treats each scalar as a single unit rather than a digit string. We call it polar because the fractional part is represented by its angle on the unit circle.

Formally, any scalar $x$ is decomposed into \textbf{a sign, $N$ integer digits, and a fractional part}:
\begin{equation}
    x = s \cdot \left(\sum\nolimits_{i=0}^{N-1}
    d_i \cdot 10^{i} + f\right)
    \label{eq:decomp}
\end{equation}
where $s \in \{-1, +1\}$, $d_i \in \{0, 1, \ldots, 9\}$ is the $i$-th integer digit, and $f \in [0,1)$ is the fractional part. We set $N$ to 9, which covers the metric coordinates produced by Eq.~\eqref{eq:geo} as well as all dynamic attributes, and pad shorter values with leading zeros.

Instead of feeding the raw digits $d_i$, we augment each integer position with the magnitude that follows it:
\begin{equation}
    \bar{d}_i = \left(|x| \,/\, 10^{i}\right) \bmod 10
    \label{eq:digit_enc}
\end{equation}
so that $\bar{d}_i \in [d_i, d_i + 1)$ carries both the digit itself and a fractional summary of all lower positions. This makes integer carries continuous: $399 \rightarrow 400$ moves $\bar{d}_2$ smoothly from $3.99$ to $4.00$ instead of changing three digits at once. Symmetrically, encoding $f$ as $(\cos 2\pi f, \sin 2\pi f)$ makes the fractional carry continuous, since $f = 0.99$ and $f = 0.01$ are adjacent on the unit circle.

Concatenating all components yields the polar vector
\begin{equation}
    \mathbf{r}(x) = \left[s,\ \bar{d}_0,\ 
    \bar{d}_1,\ \ldots,\ \bar{d}_{N-1},\ 
    \cos 2\pi f,\ \sin 2\pi f\right]^\top 
    \in \mathbb{R}^{N+3}.
    \label{eq:polar_vec}
\end{equation}

Since the entries of $\mathbf{r}(x)$ play different semantic roles, we lift each of them into a $d_e$-dim vector and add a learnable field embedding:
\begin{equation}
    \tilde{\mathbf{r}}_j(x) = r_j(x)\,\mathbf{w} + \mathbf{b}
    + \mathbf{e}_{\varphi(j)} \in \mathbb{R}^{d_e},
    \quad j = 0, \ldots, N+2
    \label{eq:field_aug}
\end{equation}
where $r_j(x)$ is the $j$-th entry of $\mathbf{r}(x)$, $\mathbf{w}, \mathbf{b} \in \mathbb{R}^{d_e}$ are shared learnable parameters, and $\varphi(j) \in \{\mathit{sign}, \mathit{int}, \mathit{frac}\}$ returns the field type of position $j$. All integer positions share a single embedding $\mathbf{e}_{\mathit{int}}$, because their order is already encoded in $\bar{d}_j$. We write $\tilde{\mathbf{r}}(x) = \mathrm{vec}\big([\tilde{\mathbf{r}}_0(x), \ldots, \tilde{\mathbf{r}}_{N+2}(x)]\big) \in \mathbb{R}^{(N+3)d_e}$ for the resulting representation. Together, the two carry-smoothing encodings and the field embeddings ensure that numerically proximate values are mapped to geometrically adjacent points in the input space.

\subsection{Numerical Encoder}
\label{sec:nce}

We pass the field-augmented polar representation $\tilde{\mathbf{r}}(x)$ through the Numerical Encoder $\mathcal{E}(\cdot)$, a multi-layer perceptron (MLP), to obtain a numerical embedding:
\begin{equation}
    \mathbf{h} = \mathcal{E}(\tilde{\mathbf{r}}(x)) 
    \in \mathbb{R}^{d}
    \label{eq:encoder}
\end{equation}
where $d$ is the encoder embedding dimension. Note that $d$ is far larger than the dimension of $\tilde{\mathbf{r}}(x)$, so $\mathcal{E}(\cdot)$ lifts a single scalar into a high-dimensional space rather than compressing it.

To make this space numerically structured, we pretrain $\mathcal{E}(\cdot)$ with a multi-task objective requiring every component of the decomposition to be recoverable from $\mathbf{h}$ alone. The sign and the $N$ integer positions are predicted by softmax classifiers,
\begin{align}
    \mathbf{p}_{sign} = 
    \text{softmax}(\mathbf{W}_{sign}\mathbf{h} + 
    \mathbf{b}_{sign}) \in \mathbb{R}^2, \\
    \mathbf{p}_{int}^{(i)} = 
    \text{softmax}(\mathbf{W}_{int}^{(i)}\mathbf{h} + 
    \mathbf{b}_{int}^{(i)}) \in \mathbb{R}^{10},
    \label{eq:cls_head}
\end{align}
for $i = 0, 1, \ldots, N-1$, while the fractional part is predicted by regression through a sigmoid-activated linear layer:
\begin{equation}
    \hat{f}(\mathbf{h}) = 
    \sigma(\mathbf{W}_f\mathbf{h} + \mathbf{b}_f) 
    \in (0,1)
    \label{eq:frac_head}
\end{equation}

This objective is not a trivial reconstruction. Since $\mathcal{E}(\cdot)$ increases dimensionality, no information is at risk of being discarded; what the objective constrains is \textit{how} that information is arranged. A single $\mathbf{h}$ must simultaneously support $N$ independent $10$-way classifications, one per digit position, together with a sign classification and a fractional regression, which forces the digit-wise magnitude structure of $x$ to be linearly decodable from separate directions of the embedding. This is also what later allows the same three heads to be attached to the LLM output (\S\ref{sec:integrate}), since they define what counts as a valid numerical representation.

For a single scalar $x$, the pretraining loss combines three heads:
\begin{equation}
    \ell_{num}(x) = \lambda_s \mathcal{L}_{sign} 
    + \lambda_i \sum\nolimits_{i=0}^{N-1} \omega_i \mathcal{L}_{int}^{(i)} 
    + \lambda_f \mathcal{L}_{frac}
    \label{eq:loss_pre}
\end{equation}
where $\mathcal{L}_{sign}$ and $\mathcal{L}_{int}^{(i)}$ are cross-entropy losses for the sign and the $i$-th integer digit, $\mathcal{L}_{frac}$ is the mean squared error of the fractional regression, and $\lambda_s, \lambda_i, \lambda_f \geq 0$ balance the three terms. The positional weight $\omega_i = 1 + 0.2i$ mildly favours higher-order digits, whose errors shift the value by larger amounts, while keeping the low-order digits supervised; a steeper weighting would let the model ignore the trailing digits, which for attributes such as ratings carry most of the discriminative signal.

After pretraining, $\mathcal{E}(\cdot)$ is frozen and reused across all downstream stages of HF-SID, providing one consistent embedding space for geographic coordinates and heterogeneous numerical attributes alike. 
\subsection{Integrating Numerical Encoder into the LLM}
\label{sec:integrate}
\subsubsection{Type Embedding}
Since the Numerical Encoder is frozen, we bridge it to the LLM with two lightweight trainable components. A linear projection first aligns its output with the LLM input space:
\begin{equation}
    \mathbf{h}^{proj}_k =
    \mathbf{W}_{proj}\mathbf{h}_k + \mathbf{b}_{proj}
    \label{eq:proj}
\end{equation}
where $\mathbf{h}_k$ is the encoder output for the $k$-th numerical token, $\mathbf{W}_{proj} \in \mathbb{R}^{d_{llm} \times d}$ and $\mathbf{b}_{proj} \in \mathbb{R}^{d_{llm}}$ are learnable parameters, and $d_{llm}$ is the input embedding dimension of the LLM.

A single shared projection, however, maps every attribute into one common numeric subspace, in which a value of $0.5$ is indistinguishable whether it denotes a rating or a visit count. We therefore add a learnable type embedding, drawn from a matrix $\mathbf{M}_{\mathcal{T}} \in \mathbb{R}^{|\mathcal{T}| \times d_{llm}}$ indexed by a finite set of attribute types $\mathcal{T}$, to each numerical embedding according to its own type $\tau_k \in \mathcal{T}$:
\begin{equation}
    \hat{\mathbf{h}}_k = \mathbf{h}^{proj}_k +
    \mathbf{M}_{\mathcal{T}}[\tau_k]
    \label{eq:type_emb}
\end{equation}
Because the offset is constant within a type, magnitude relations among values of the same attribute are preserved exactly, while different attributes are translated into separate regions of the input space. This is exactly the behavior we want: the model can compare two ratings on a common scale, yet it can no longer confuse a rating with a visit count. The mechanism is analogous to positional embeddings in Transformers~\cite{Attention}, which likewise inject a prior about the role of a position without modifying its content.


\subsubsection{Fused Embedding Construction}
When a POI is serialized, every numerical value is replaced by a single placeholder token \texttt{[NUM]}, so that one scalar occupies exactly one position irrespective of how many digits it has; as a side effect this also shortens the input sequence, since a multi-digit number no longer expands into several sub-tokens. We then substitute the type-aware numerical embeddings at these positions and keep ordinary token embeddings elsewhere, before feeding the sequence to the pretrained LLM:
\begin{equation}
    \mathbf{E}_{fused}[k] =
    \begin{cases}
        \hat{\mathbf{h}}_k & \text{if } k \in \mathcal{I}_{num} \\
        \mathrm{Emb}(w_k) & \text{otherwise}
    \end{cases}
    \quad
    \mathbf{E}_{out} = \text{LLM}(\mathbf{E}_{fused})
    \label{eq:llm_out}
\end{equation}
where $\mathcal{I}_{num}$ denotes the \texttt{[NUM]} positions of the input sequence, $\mathrm{Emb}(\cdot)$ is the LLM token embedding lookup, $w_k$ is the $k$-th text token, and $\mathbf{E}_{fused}, \mathbf{E}_{out} \in \mathbb{R}^{T \times d_{llm}}$ with $T$ the total sequence length. Following common decoder-based embedding models~\cite{qwen3embedding}, we take the hidden state of the last token as the POI representation $\mathbf{E}_{out}(p) \in \mathbb{R}^{d_{llm}}$, serving as the input to HF-SID generation (\S\ref{sec:sid}).

\subsubsection{Two-Stage Continual Pre-Training}

\paragraph{Training Data Construction.} We construct two categories of training samples, one for geographic reasoning and one for numerical reasoning, to explicitly supervise the LLM on both dimensions. In every sample, all numerical quantities appear as \texttt{[NUM]} tokens and are encoded by the frozen Numerical Encoder, both in the input and in the target sequence.

\textbf{Geographic-enhanced Continual Pre-Training (Geo-CPT).} To train the model to reason about spatial relationships between POIs, we construct two types of geographic samples: \textbf{Pairwise Distance Computation.} Given two POIs $p_A$ and $p_B$, whose Cartesian coordinates $(x_A, y_A, z_A)$ and $(x_B, y_B, z_B)$ are supplied as numerical tokens, the model is asked to predict the distance between them at a single \texttt{[NUM]} position of the target sequence. The ground truth is the great-circle distance $d(p_A, p_B)$ in kilometers, computed by the Haversine formula. Note that the quantity directly available from the Cartesian inputs is the chord length $2R\sin\!\left(d/2R\right)$ rather than $d$ itself; the two agree to within $10^{-3}\%$ for POI pairs closer than $100$\,km, so at the scale relevant to LBS retrieval this target is effectively the Euclidean distance in the transformed space, while remaining the geographically meaningful quantity. \textbf{Nearest Pair Identification.} Given three POIs $p_A$, $p_B$, and $p_C$, the model is asked to identify the closest pair among $(p_A, p_B)$, $(p_A, p_C)$, and $(p_B, p_C)$, and to emit it as text tokens. This requires jointly reasoning over three pairwise distances rather than computing a single one.



\textbf{Numerical-enhanced Continual Pre-Training (Num-CPT).} We construct samples in which two POIs $p_A$ and $p_B$ are described with multiple numerical attributes, and the model is asked which of them has the larger value of a queried attribute $\tau$. The target is emitted as text tokens, and we define it over the two candidate labels $\{\mathtt{A}, \mathtt{B}\}$:
\begin{equation}
    y^{(\tau)} =
    \begin{cases}
        \mathtt{A} & \text{if } v_A^{(\tau)} > v_B^{(\tau)} \\
        \mathtt{B} & \text{if } v_A^{(\tau)} < v_B^{(\tau)}
    \end{cases}
    \label{eq:comp}
\end{equation}
where $v_A^{(\tau)}$ and $v_B^{(\tau)}$ are the values of the two POIs under attribute type $\tau$. Pairs with $v_A^{(\tau)} = v_B^{(\tau)}$ are discarded, since ties carry no comparative signal and would otherwise train a spurious preference for one of the two positions; we also balance the two labels so that the answer cannot be inferred from position alone. Instantiated across attributes, this formulation forces the model to locate the value belonging to the queried attribute and to compare it against its counterpart of the same type, even though different types differ by orders of magnitude in scale.

\paragraph{Hybrid Training Objective}

Because the target sequences contain both discrete text tokens and numerical tokens, we adopt a hybrid objective that applies a different loss at each kind of position. Let $\mathcal{I}_{text}$ and $\mathcal{I}_{num}$ denote the text and numerical positions of the target sequence, and let $\mathbf{E}_{<k}$ denote the fused prefix of position $k$, which may contain both text token embeddings and injected numerical embeddings. At text positions we apply the standard cross-entropy next-token prediction (NTP) loss:
\begin{equation} 
    \mathcal{L}_{CE} = 
    -\frac{1}{|\mathcal{I}_{text}|}\sum\nolimits_{k \in \mathcal{I}_{text}}
    \log P\left(w_k \mid \mathbf{E}_{<k}\right)
    \label{eq:loss_ce}
\end{equation}
where $w_k$ is the ground-truth text token at position $k$. Conditioning on $\mathbf{E}_{<k}$ rather than on $w_{<k}$ is what allows a numerical value to act as context for the tokens that follow it.

At numerical positions we reuse the multi-task loss $\ell_{num}$, applying the same three decoder heads to the LLM output representation:
\begin{equation}
    \mathcal{L}_{num}^{NTP} =
    \frac{1}{|\mathcal{I}_{num}|}\sum\nolimits_{k \in \mathcal{I}_{num}} \ell_{num}(x_k)
    \label{eq:loss_num_ntp}
\end{equation}
where $x_k$ is the target value at position $k$, and its sign, integer, and fractional losses are computed from the LLM hidden state at that position. Both terms are averaged over their own positions so that their relative weight does not drift with the text-to-number ratio of a sample. And each of the two stages is trained with
\begin{equation}
    \mathcal{L}_{NTP} = \mathcal{L}_{CE} +
    \alpha \cdot \mathcal{L}_{num}^{NTP}
    \label{eq:loss_ntp}
\end{equation}
where $\alpha \geq 0$ balances the numerical alignment loss against next-token prediction. We run Geo-CPT first and Num-CPT second, since geographic fidelity is the dominant factor for LBS retrieval and therefore benefits from being established on the unmodified backbone.

\subsection{HF-SID Generation with Structure-based Contrastive Learning}
\label{sec:sid}

We obtain Semantic IDs by residual-quantizing the POI representations of the entire corpus with three layers of K-means. Writing $\mathbf{R}_p^{(0)} = \mathbf{E}_{out}(p) \in \mathbb{R}^{d_{llm}}$, layer $l$ fits a codebook $\mathcal{C}^{(l)} = \{\mathbf{c}_1^{(l)}, \ldots, \mathbf{c}_{N_c}^{(l)}\}$ of size $N_c$ over the $l$-th order residuals of all POIs, assigns each POI to its nearest centroid, and forwards the remaining residual:
\begin{equation}
    s_p^{l} = \arg\min_{j} \big\lVert \mathbf{R}_p^{(l-1)} 
    - \mathbf{c}_j^{(l)} \big\rVert_2,
    \qquad
    \mathbf{R}_p^{(l)} = \mathbf{R}_p^{(l-1)} 
    - \mathbf{c}_{s_p^{l}}^{(l)},
    \label{eq:sid_coarse}
\end{equation}
where $\mathbf{E}_{out}(p)$ is the last-token embedding of the LLM and every layer uses the same codebook size $N_c$. The first two layers are applied in standard form and yield the coarse tokens $s_p^1$ and $s_p^2$, which reflect the geographic and numerical semantics.

Precisely because these two layers are geographically compact, POIs located in the same business district are largely collapsed into a shared prefix $\langle s_p^1, s_p^2 \rangle$ irrespective of what they actually are. We therefore refine the second-order residual $\mathbf{R}_p^{(2)}$ before the final quantization, supervised by the two-level tag of each POI. Two POIs form a positive pair only when both levels of their tags coincide, and every other POI in the mini-batch serves as a negative, so that POIs agreeing on the coarse level alone act as hard negatives. The residual is linearly projected,
\begin{equation}
    \mathbf{z}_p = \mathbf{W}_{s}\,\mathbf{R}_p^{(2)} 
    + \mathbf{b}_{s}
    \label{eq:tag_proj}
\end{equation}
and optimized with an InfoNCE objective over inner products:
\begin{equation}
    \mathcal{L}_{s}(p_i, p_j) = -\log 
    \frac{\exp\left(\mathbf{z}_{p_i}^\top 
    \mathbf{z}_{p_j} / \gamma\right)}
    {\exp\left(\mathbf{z}_{p_i}^\top 
    \mathbf{z}_{p_j} / \gamma\right) + 
    \sum_{p_k \in \mathcal{B}^{-}} 
    \exp\left(\mathbf{z}_{p_i}^\top
    \mathbf{z}_{p_k} / \gamma\right)}
    \label{eq:loss_tag}
\end{equation}
where $\mathbf{W}_{s} \in \mathbb{R}^{d_z \times d_{llm}}$ and $\mathbf{b}_{s} \in \mathbb{R}^{d_z}$ are learnable projection parameters with output dimension $d_z$ and the only parameters updated at this stage, $\mathcal{B}^{-}$ collects the in-batch POIs whose tag differs from that of $p_i$ at either level, $\gamma > 0$ is a temperature hyperparameter, and the loss is averaged over all positive pairs in the batch. Since the LLM and the first two codebooks stay fixed, $\mathbf{R}_p^{(2)}$ is a constant input and the refinement never invalidates the coarse tokens.

The refined residual is then quantized by the third K-means layer into $s_p^3$, so that POIs sharing a fine-grained tag tend to fall into the same third-layer cluster, and each POI is finally represented as
\begin{equation}
    \mathbf{HF-SID}_p = \left\langle s_p^1,\ 
    s_p^2,\ s_p^3 \right\rangle
    \label{eq:sid}
\end{equation}
Note that this last layer operates in the projected space $\mathbb{R}^{d_z}$ rather than in $\mathbb{R}^{d_{llm}}$, and therefore no longer reconstructs $\mathbf{R}_p^{(2)}$ in the strict residual-quantization sense. This is deliberate: an SID serves only as an identifier to be generated autoregressively and is never used to reconstruct the POI representation, so the final layer can trade reconstruction error for discriminative power. Restricting the projection to the last layer is what keeps the geographic and numerical fidelity of $\langle s_p^1, s_p^2 \rangle$ intact, as the analysis in \S\ref{sec:analysis} confirms.

The resulting sequence retains the prefix-sharing property of residual quantization: POIs with similar representations agree on their leading tokens, while the last token separates them by fine-grained tag. Candidate generation can thus be performed by beam search constrained to the prefix tree of valid SIDs.

\section{Experiments}

\subsection{Experimental Setup}

\textbf{Datasets.} We evaluate on two industrial datasets from the AMap  platform\footnote{\url{https://amap.com}} and two public Foursquare check-in  datasets (NYC, TKY). \textbf{AMap-L} is built from one full day of anonymized  user interaction logs. For data quality, we remove POIs with no associated  interaction sequence and exclude sessions without authenticated user identifiers.  \textbf{AMap-S} is a  publicly released subset from the same platform: smaller in scale but retaining  the same attributes, it can serve as a standardized benchmark for  future work. Statistics  are summarized in Table~\ref{tab:dataset_stats}.

\textbf{Baselines.} We compare our HF-SID against the following SID construction methods. RQ-KMeans~\cite{rqkmeans}, RQ-OPQ~\cite{chen2025onesearch}, GNPR-SID~\cite{GNPR-SID}, Cosine-RQ~\cite{zhou2026hymirec}, Pro-GEO~\cite{ProGEO}, GenPOI~\cite{GENPOI}, GeoGR~\cite{GeoGR}. Detailed information can be found in the given references. For fair comparison, all baselines are re-implemented under the same configuration as HF-SID: codebook depth $L=3$, per-layer codebook size $N_c=4096$ on AMap-L and $N_c=512$ on AMap-S, and identical dataset splits. 


\textbf{Evaluation Metrics.} We evaluate both the quality of HF-SID and downstream performance through Next POI recommendation task~\cite{ProGEO, GNPR-SID}. \textbf{(1) Quantization Metrics of SID.}
Follow prior work~\cite{ProGEO}, we use 4 metrics: ICR, Avg.Dist, p90\,Dist, p95\,Dist. ICR (Independent Codeword Rate, $\uparrow$) measures the degree to which codewords are assigned uniquely across items. Avg.Dist ($\downarrow$) computes the mean Euclidean distance from each POI’s position to the centroid of its quantification SID, providing a measure of intra-group spatial compactness. Additionally, we report p90 Dist and p95 Dist, representing the maximum distance to the SID centroid within the top 90\% and 95\% of POIs. \textbf{(2) Offline Recommendation Metrics.}
Hit@200 and Hit@300 ($\uparrow$) measure the proportion of ground-truth target POIs appearing within the top-200 and top-300 generated candidates, respectively, jointly reflecting retrieval precision and recall coverage. \textbf{(3) Online Recommendation Metrics.} We evaluate online performance using five metrics: WinRate ($\uparrow$), a composite metric that quantifies overall user engagement gains by weighting strategic actions such as clicks, route navigation, and bookings according to their value, reflecting holistic improvements in both user satisfaction and platform utility. PV\_CTR and UV\_CTR measure click-through rates at the exposure and user levels, respectively. PV\_CVR and UV\_CVR measure conversion rates at the exposure and user levels, respectively.

\begin{table}[t]
    \setlength{\abovecaptionskip}{0.1cm}
    \setlength{\belowcaptionskip}{-0.2cm}
\centering
    \caption{Dataset statistics.}
    \label{tab:dataset_stats}
    \setlength{\tabcolsep}{4.8mm}
    \scalebox{0.80}{
        \begin{tabular}{ccccc}
        \hline
        \hline
        Dataset & \#Users & \#POIs & \#Inter. & Avg.Len \\
        \hline
        NYC    & 1,083 & 5,135  & 104,074 & 136 \\
        TKY    & 2,293 & 7,873  & 361,430 & 195 \\
        AMap-L & 11.0M & 48.1M  & 16.3M   & 305 \\
        AMap-S & 0.90M & 4.58M  & 0.97M   & 453 \\
        \hline
        \hline
        \end{tabular}
    }
\end{table}

\begin{table*}[t]
  \setlength{\abovecaptionskip}{0.2cm}
    \setlength{\belowcaptionskip}{-0.4cm}
\centering
\caption{
Performance comparison on AMap-L and AMap-S datasets. \textbf{Bold} denotes the best result within each dataset.
}
\label{tab:main_amap}
\setlength{\tabcolsep}{4.3mm}
\scalebox{0.85}{
    \begin{tabular}{l|l|c|cccc|cc}
    \hline
    \hline
    \multirow{2}{*}{Dataset}& \multirow{2}{*}{Method}& \multirow{2}{*}{SID Length}& \multicolumn{4}{c|}{Quantization Metrics}& \multicolumn{2}{c}{Rec. Metrics} \\
    \cline{4-9}
    &&& ICR($\uparrow$)& Avg.Dist.($\downarrow$)& p90 Dist.($\downarrow$)& p95 Dist.($\downarrow$)& Hit@200($\uparrow$)& Hit@300($\uparrow$) \\
    \hline
    \multirow{8}{*}{AMap-L}& RQ-KMeans~\cite{rqkmeans}& 3& 80.92\% & 6.07\,km & 2.66\,km & 15.21\,km& 54.47\% & 56.02\% \\

    & RQ-OPQ~\cite{chen2025onesearch}& 3& 98.93\% & 7.27\,km & 5.03\,km & 23.78\,km& 54.88\% & 55.31\% \\

    & GNPR-SID~\cite{GNPR-SID}& 3& 77.57\% & 10.67\,km & 5.08\,km & 16.27\,km& 58.38\% & 67.65\% \\

    & Cosine-RQ~\cite{zhou2026hymirec}& 3& 82.63\% & 5.26\,km & 2.33\,km & 13.32\,km& 55.62\% & 58.83\% \\
    
    & Pro-GEO~\cite{ProGEO}& 3& 81.18\% & 7.02\,km & 4.84\,km & 23.33\,km& 65.69\% & 70.51\% \\

    & GeoGR~\cite{GeoGR}& 3& 54.69\% & 0.49\,km & 0.74\,km & 1.19\,km& 77.58\% & 80.42\% \\
    
    & GenPOI~\cite{GENPOI}& 6(3+3)& 83.80\% & 0.97\,km & 0.55\,km & 5.53\,km& 70.61\% & 73.82\% \\
    
    & \textbf{HF-SID (Ours)}& 3& \textbf{84.22\%}& \textbf{0.25\,km}& \textbf{0.12\,km}& \textbf{0.69\,km}& \textbf{81.24\%}& \textbf{83.36\%} \\
    
    \hline
    
    \multirow{9}{*}{AMap-S}& RQ-KMeans~\cite{rqkmeans}& 3& 61.34\% & 37.91\,km & 88.48\,km & 207.85\,km& 27.36\% & 33.07\% \\

    & RQ-OPQ~\cite{chen2025onesearch}& 3& 98.71\% & 41.20\,km & 96.94\,km & 226.36\,km& 26.26\% & 26.71\% \\

    & GNPR-SID~\cite{GNPR-SID}& 3& 47.52\% & 77.82\,km & 229.70\,km & 485.91\,km& 29.64\% & 36.83\% \\

    & Cosine-RQ~\cite{zhou2026hymirec}& 3& 63.68\% & 36.32\,km & 74.59\,km & 197.81\,km& 28.95\% & 33.78\% \\
    
    & Pro-GEO~\cite{ProGEO}& 3& 64.64\% & 24.25\,km & 49.54\,km & 132.85\,km& 34.54\% & 38.39\% \\
    
    & GeoGR~\cite{GeoGR}& 3& 37.83\% & 8.94\,km & 14.07\,km & 46.15\,km& 35.02\% & 40.50\% \\
    
    & GenPOI~\cite{GENPOI}& 5(2+3)& 62.12\% & 4.84\,km & 15.71\,km & 25.14\,km& 38.04\% & 43.29\% \\
    
    & GenPOI~\cite{GENPOI}& 6(3+3)& \textbf{65.97\%} & \textbf{3.22\,km} & 11.30\,km & 18.19\,km& 42.20\% & 46.91\% \\
    
    & \textbf{HF-SID (Ours)}     & 3     & 59.79\%     & 3.99\,km     & \textbf{5.49\,km}     & \textbf{11.12\,km}     & \textbf{48.98\%}     & \textbf{57.92\%} \\
    
    \hline
    \hline
\end{tabular}
}
\end{table*}

\textbf{Implementation Details.} All experiments are implemented in PyTorch. Following prior work~\cite{geonum}, we implement the Numerical Encoder. The embedding model's backbone is initialized from Qwen3-0.6B~\cite{qwen3embedding} for POI representation learning and SID generation, while the downstream next-POI prediction task uses Qwen3-4B~\cite{qwen3} with full-parameter supervised fine-tuning (SFT). Beam search is performed with a width of 300 to generate the final list of POIs. Continual pre-training runs for 1 epoch with AdamW~\cite{loshchilov2019decoupledweightdecayregularization} at learning rate $1\times10^{-3}$. SFT also runs for 1 epoch with a cosine scheduler, learning rate $2\times10^{-5}$, per-device batch size 8, and gradient accumulation of 4 steps, using DeepSpeed ZeRO-2~\cite{rajbhandari2020zeromemoryoptimizationstraining} with bf16 mixed precision. For RQ-KMeans we set $L=3$ layers, consistent with prior work~\cite{deng2025onerec}. All experiments run on 64 NVIDIA A100 GPUs.

\subsection{Overall Performance}
\label{sec:main_results}
We examine whether the representation-level fidelity that HF-SID enforces actually translates into better SIDs and better retrieval performance. Table~\ref{tab:main_amap} compares HF-SID against seven baselines covering the three ways of building POI SIDs: text-only quantization, post-hoc geographic injection, and embedding-space alignment.

\textbf{HF-SID Quality.}
On both datasets the compactness of the SID space improves monotonically with how early geography enters the pipeline, and HF-SID, which intervenes earliest, is the most compact of all. Methods that quantize text-only embeddings stay at $5.26$--$10.67$\,km of Avg.\,Dist. on AMap-L and $36.32$--$77.82$\,km on AMap-S, regardless of how their codebooks are refined: orthogonal product quantization~\cite{chen2025onesearch}, a codeword diversity loss~\cite{GNPR-SID}, and stronger text representations~\cite{zhou2026hymirec} all leave the geographic error essentially untouched, because none of them puts geographic information into the embedding to begin with. Injecting geography afterwards helps ($7.02$\,km for Pro-GEO, $0.97$\,km for the 6-token GenPOI), and reshaping the embedding space with behavioral supervision helps further ($0.49$\,km for GeoGR), but HF-SID reaches $0.25$\,km and $3.99$\,km, cutting Avg.\,Dist. by $95.9\%$ and $89.5\%$ relative to plain RQ-KMeans and by a further $48.98\%$ and $55.4\%$ relative to GeoGR. Its advantage is largest in the tail, which is the regime that produces visibly wrong candidates: p90 Dist. drops by $83.8\%$ and $61.0\%$ and p95 Dist. by $42.0\%$ and $75.9\%$ against the same baseline. Notably, on AMap-S HF-SID halves the p90 Dist. of the 6-token GenPOI ($11.30 \rightarrow 5.49$\,km) and reduces its p95 Dist. by $38.9\%$ ($18.19 \rightarrow 11.12$\,km) using half the identifier length, although GenPOI retains a slightly lower mean of $3.22$\,km under that doubled token budget.

Two baselines further show that codeword uniqueness on its own is not a meaningful objective. RQ-OPQ attains a near-saturated ICR of $98.93\%$ and $98.71\%$, yet its Avg.\,Dist. is worse than that of plain RQ-KMeans on both datasets and its Hit@300 is the lowest of all methods on both ($55.31\%$ and $26.71\%$): spreading items evenly over the codebook says nothing about whether the code carries usable structure. GeoGR fails in the opposite direction, buying geographic compactness with the lowest ICR of the table ($54.69\%$ and $37.83\%$), because contrastive fine-tuning on co-visit pairs collapses behaviorally similar POIs onto shared codewords. HF-SID avoids both failure modes because it encodes the coordinate itself rather than a proxy of it, and with an ICR of $84.22\%$ on AMap-L it is the most collision-resistant among all methods whose Avg.\,Dist. stays below $1$\,km.

\textbf{Retrieval Performance.}
The gains in SID quality carry over to downstream retrieval in full, and the ordering of methods by Hit mirrors their ordering by SID compactness almost exactly. All methods are fine-tuned under identical settings and differ only in their SIDs. HF-SID attains Hit@200 of $81.24\%$ and $48.98\%$ and Hit@300 of $83.36\%$ and $57.92\%$, exceeding the strongest baseline on each dataset by $3.66$ and $6.78$ points on Hit@200 and by $2.94$ and $11.01$ points on Hit@300, the latter amounting to a $23.48\%$ relative gain on AMap-S while using half the tokens of the 6-token GenPOI it surpasses. Across the eight AMap-L systems, Hit@200 rises from the $54$--$58\%$ range for the four text-only methods, to $65.69\%$ and $70.61\%$ once geography is injected post hoc, to $77.58\%$ and $81.24\%$ once the embedding itself becomes geography-aware. That the two orderings coincide is the central evidence of this paper: the retrieval gain is not an artefact of a particular codebook design but a direct consequence of how much geographic and numerical fidelity survives into the identifier.

\begin{table}[t]
  \setlength{\abovecaptionskip}{0.2cm}
    \setlength{\belowcaptionskip}{-0.4cm}
\centering
\caption{
    Ablation study on AMap-L.
}
\label{tab:ablation}
\setlength{\tabcolsep}{2.0mm}
\scalebox{0.83}{
    \begin{tabular}{l | cc | cc}
    \hline
    \hline
    Method
    & Hit@200 ($\uparrow$) & $\Delta$Hit@200 & Hit@300 ($\uparrow$) & $\Delta$Hit@300 \\
    \hline
    \textbf{HF-SID (Full)}
    & \textbf{81.24\%}& --  & \textbf{83.36\%}  & -- \\
    \hline
    w/o Num
    & 77.73\% & $-$4.32\% & 80.65\%  & $-$3.25\% \\
    w/o Struct
    & 71.88\%  & $-$11.52\% & 78.35\% & $-$6.01\% \\
    w/o Geo
    & 68.34\%  & $-$15.88\% & 71.11\% & $-$14.69\% \\
    \hline
    w/o All (Baseline)
    & 54.47\%  & $-$32.95\% & 56.02\% & $-$32.79\% \\
\hline
\hline
\end{tabular}
}
\end{table}

\subsection{Ablation Study}
\label{sec:ablation}
We ablate HF-SID on AMap-L in a leave-one-out fashion. No variant removes any \emph{information}: when a component is disabled, the corresponding attribute stays in the POI description as plain text. \textbf{w/o Geo} drops the Cartesian transformation, the Numerical Encoder on coordinates, and Geo-CPT; \textbf{w/o Num} drops the Numerical Encoder on dynamic attributes, the Type Embedding, and Num-CPT; \textbf{w/o Struct} drops the contrastive refinement, letting the third layer quantize $\mathbf{R}_p^{(2)}$ directly; \textbf{w/o All} reduces to plain RQ-KMeans. All variants therefore see identical content and differ only in how faithfully it is represented, so the drops below measure the value of representation fidelity rather than of information availability.

\textbf{Every component contributes, and the size of its contribution follows the level of the SID hierarchy at which it acts.}
Removing the geographic pathway costs the most, with Hit@200 falling from $81.24\%$ to $68.34\%$ ($-15.88\%$) and Hit@300 from $83.36\%$ to $71.11\%$ ($-14.69\%$): geographic fidelity is established before quantization and thus shapes the first two codebook layers, that is, the entire prefix tree over which generation is constrained, so degrading it corrupts every subsequent decoding decision. Structure-based Contrastive Learning acts on only one token out of three, and its cost accordingly halves from Hit@200 ($-11.52\%$) to Hit@300 ($-6.01\%$), indicating that tag purity matters mainly for ranking the correct POI early rather than for reaching it at all. The numerical pathway is the smallest ($-4.32\%$ and $-3.25\%$), as ratings and visit counts refine the ordering among already plausible candidates instead of deciding which region of the identifier space is explored.

\textbf{The entire gap to the baseline is attributable to the representation alone.}
With all three components disabled the model returns to plain RQ-KMeans at $54.47\%$ and $56.02\%$, so HF-SID's $26.77$-point gain on Hit@200 is obtained without changing the quantization algorithm, the codebook size, or the identifier length. The same procedure that ranks last in Table~\ref{tab:main_amap} ranks first once the representation it quantizes is made geography- and numeracy-aware.

\subsection{Visualization Analysis of HF-SID}
\label{sec:analysis}

Through detailed experiments, we explore how much geographic, numerical, and structural information HF-SID carries.

\subsubsection{Visualization of Geographic Information}
\begin{figure}[t]
    \setlength{\abovecaptionskip}{0.4cm}
    \setlength{\belowcaptionskip}{-0.2cm}
    \centering
    \begin{subfigure}[t]{0.30\linewidth}
        \centering
        \includegraphics[width=\linewidth]{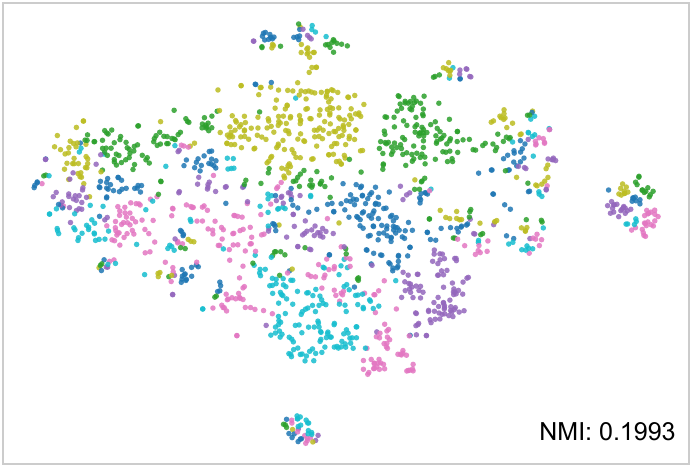}
        \caption{Province Baseline}
        \label{fig:province_baseline}
    \end{subfigure}
    \hfill
    \begin{subfigure}[t]{0.30\linewidth}
        \centering
        \includegraphics[width=\linewidth]{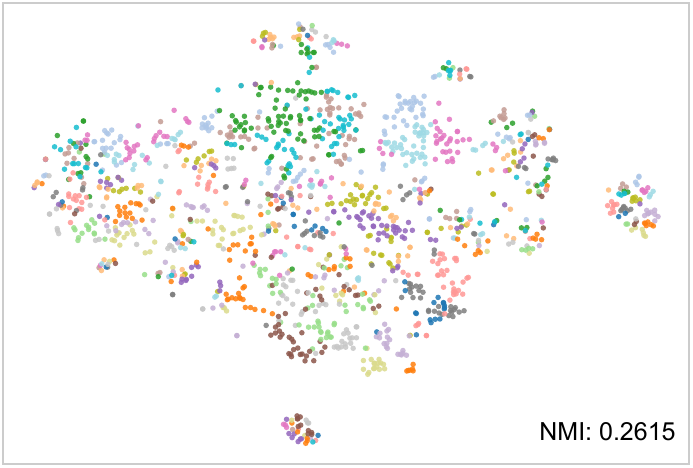}
        \caption{City Baseline}
        \label{fig:city_baseline}
    \end{subfigure}
    \hfill
    \begin{subfigure}[t]{0.30\linewidth}
        \centering
        \includegraphics[width=\linewidth]{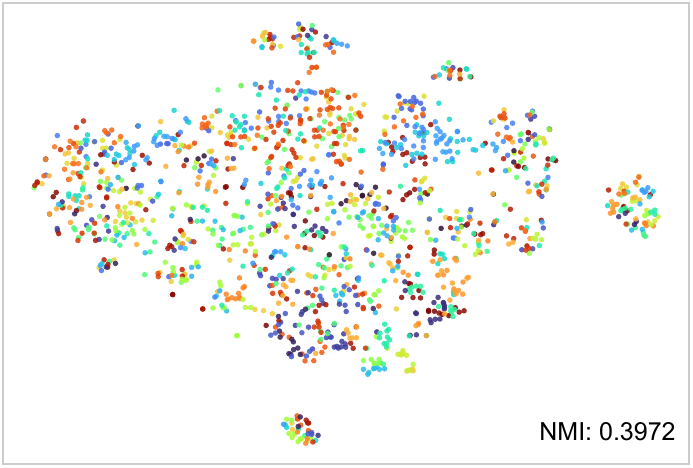}
        \caption{District Baseline}
        \label{fig:district_baseline}
    \end{subfigure}
    
    \vspace{0.2cm}

    \begin{subfigure}[t]{0.30\linewidth}
        \centering
        \includegraphics[width=\linewidth]{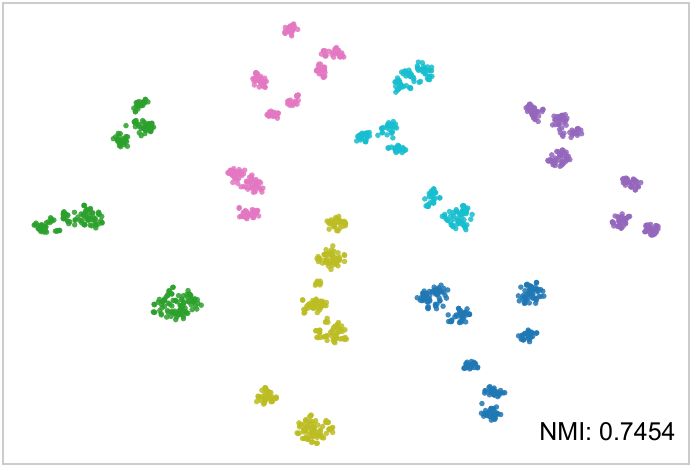}
        \caption{Province HF-SID}
        \label{fig:province_HF-SID}
    \end{subfigure}
    \hfill
    \begin{subfigure}[t]{0.30\linewidth}
        \centering
        \includegraphics[width=\linewidth]{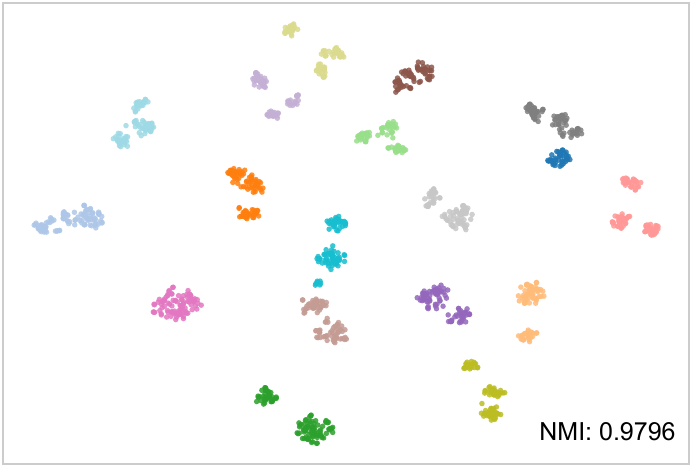}
        \caption{City HF-SID}
        \label{fig:city_HF-SID}
    \end{subfigure}
    \hfill
    \begin{subfigure}[t]{0.30\linewidth}
        \centering
        \includegraphics[width=\linewidth]{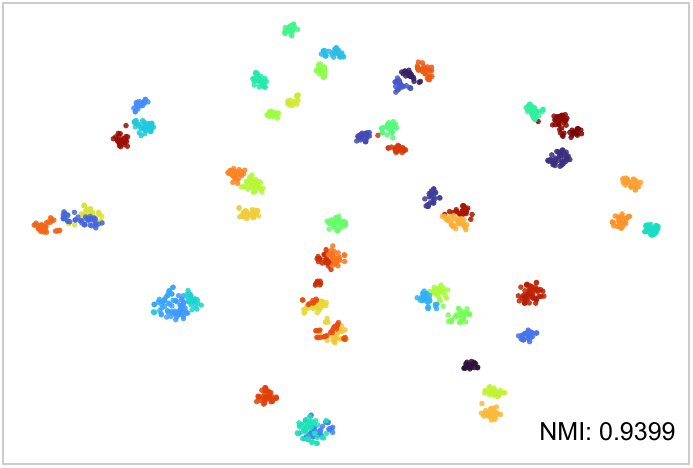}
        \caption{District HF-SID}
        \label{fig:district_HF-SID}
    \end{subfigure}
    \caption{t-SNE visualization of POI embeddings.}
    \label{fig:comparison}
\end{figure}

We sample $1{,}554$ POIs across three administrative levels (Province, City, and District), embed them with t-SNE using the geographic label as color, and report the NMI before and after Geo-CPT. As shown in Figure~\ref{fig:comparison}, the baseline embeddings remain largely overlapped at every granularity, whereas HF-SID forms tight and well-separated clusters, with NMI rising from $0.1993$ to $0.7454$ at Province, $0.2615$ to $0.9796$ at City, and $0.3972$ to $0.9399$ at District. This embedding-space geometry is exactly what produces the SID-space compactness of Table~\ref{tab:main_amap}: fidelity established in the representation survives into the code.

\begin{table*}[t]
    \setlength{\abovecaptionskip}{0.1cm}
    \setlength{\belowcaptionskip}{-0.2cm}
\centering
\caption{
    Intra-SID numerical attribute dispersion comparison across methods. \textbf{Bold} denotes the best result.
}
\label{tab:numerical_dispersion}
\setlength{\tabcolsep}{1.8mm}
\scalebox{0.73}{
\begin{tabular}{l|ccc|ccc|ccc|ccc|ccc}
\hline
\hline
\multirow{2}{*}{\textbf{Method}}
    & \multicolumn{3}{c|}{\textbf{Rating} $\downarrow$}
    & \multicolumn{3}{c|}{\textbf{POI Visited} $\downarrow$}
    & \multicolumn{3}{c|}{\textbf{POI Index} $\downarrow$}
    & \multicolumn{3}{c|}{\textbf{Collect UV} $\downarrow$}
    & \multicolumn{3}{c}{\textbf{Week View UV} $\downarrow$} \\
    & Avg.Std & Avg.CV & Avg.MAD
    & Avg.Std & Avg.CV & Avg.MAD
    & Avg.Std & Avg.CV & Avg.MAD
    & Avg.Std & Avg.CV & Avg.MAD
    & Avg.Std & Avg.CV & Avg.MAD \\
\hline
GeoGR
    & 0.2474 & 0.1001 & 0.2880
    & $1.92\times10^{11}$ & 1.2474 & $1.92\times10^{11}$
    & 0.6121 & 0.4063 & 0.5695
    & 24.953 & 0.4537 & 24.897
    & 7.631  & 0.7137 & 7.196  \\
Pro-GEO
    & 0.1831 & 0.0918 & 0.2610
    & $4.39\times10^{11}$ & 1.0661 & $4.39\times10^{11}$
    & 0.4894 & 0.3381 & 0.4696
    & 30.018 & 0.3716 & 33.077
    & 7.924  & 0.6292 & 7.830  \\
GenPOI
    & 0.2027 & 0.0947 & 0.2731
    & 253.178 & 1.0600 & 236.968
    & 0.5133 & 0.3462 & 0.4951
    & 30.554 & 0.3701 & 33.572
    & 8.188  & 0.6244 & 8.166  \\
RQ-KMeans
    & 0.2138 & 0.0982 & 0.2808
    & 222.486 & 1.0753 & 208.249
    & 0.5124 & 0.3554 & 0.4934
    & 27.755 & 0.3697 & 30.237
    & 7.573  & 0.6307 & 7.515  \\
\hline
HF-SID 
    & \textbf{0.1817} & \textbf{0.0916} & \textbf{0.2608}
    & \textbf{49.955} & \textbf{1.0297} & \textbf{48.180}
    & \textbf{0.4438} & \textbf{0.3138} & \textbf{0.4251}
    & \textbf{15.426} & \textbf{0.3650} & \textbf{17.545}
    & \textbf{4.366}  & \textbf{0.5962} & \textbf{4.451}  \\
\hline
\hline
\end{tabular}
}
\end{table*}
\subsubsection{Analysis of Numerical Awareness}

To assess numerical awareness, we measure the intra-HF-SID dispersion of five heterogeneous attributes: Rating, POI Visited, POI Index, Collect UV, and Week View UV. For every HF-SID cluster with more than one POI we compute the Standard Deviation (Std), Coefficient of Variation (CV), and Mean Absolute Deviation (MAD), and report their macro-averages, where lower values indicate greater numerical coherence (Table~\ref{tab:numerical_dispersion}).

HF-SID attains the best dispersion on all five attributes, and the margin scales with how demanding the attribute is. On Rating, which is confined to a narrow $0$--$5$ range, the improvement over the second-best baseline Pro-GEO is small but consistent (Avg.Std $0.1831 \rightarrow 0.1817$, Avg.CV $0.0918 \rightarrow 0.0916$, Avg.MAD $0.2610 \rightarrow 0.2608$). The advantage widens sharply on heavy-tailed attributes spanning several orders of magnitude: on POI Visited, HF-SID cuts Avg.Std by $77.5\%$ ($222.486 \rightarrow 49.955$) and Avg.MAD by $76.9\%$ ($208.249 \rightarrow 48.180$) against the second-best baseline on each metric, and also lowers Avg.CV from $1.0600$ to $1.0297$, showing that relative dispersion is reduced even under extreme scale. The same pattern holds on POI Index (Avg.Std $-9.3\%$, Avg.CV $-7.2\%$, Avg.MAD $-9.5\%$ over Pro-GEO) and on Collect UV and Week View UV (Avg.Std $-38.2\%$ and $-42.4\%$, Avg.MAD $-29.5\%$ and $-38.2\%$ over their second-best baselines). That a single frozen encoder handles all five distributions confirms that the numerical pathway generalizes across scales rather than fitting any one attribute.

\begin{figure*}[t]
    \setlength{\abovecaptionskip}{0.1cm}
    \setlength{\belowcaptionskip}{-0.2cm}
    \centering
    \includegraphics[width=0.98\linewidth]{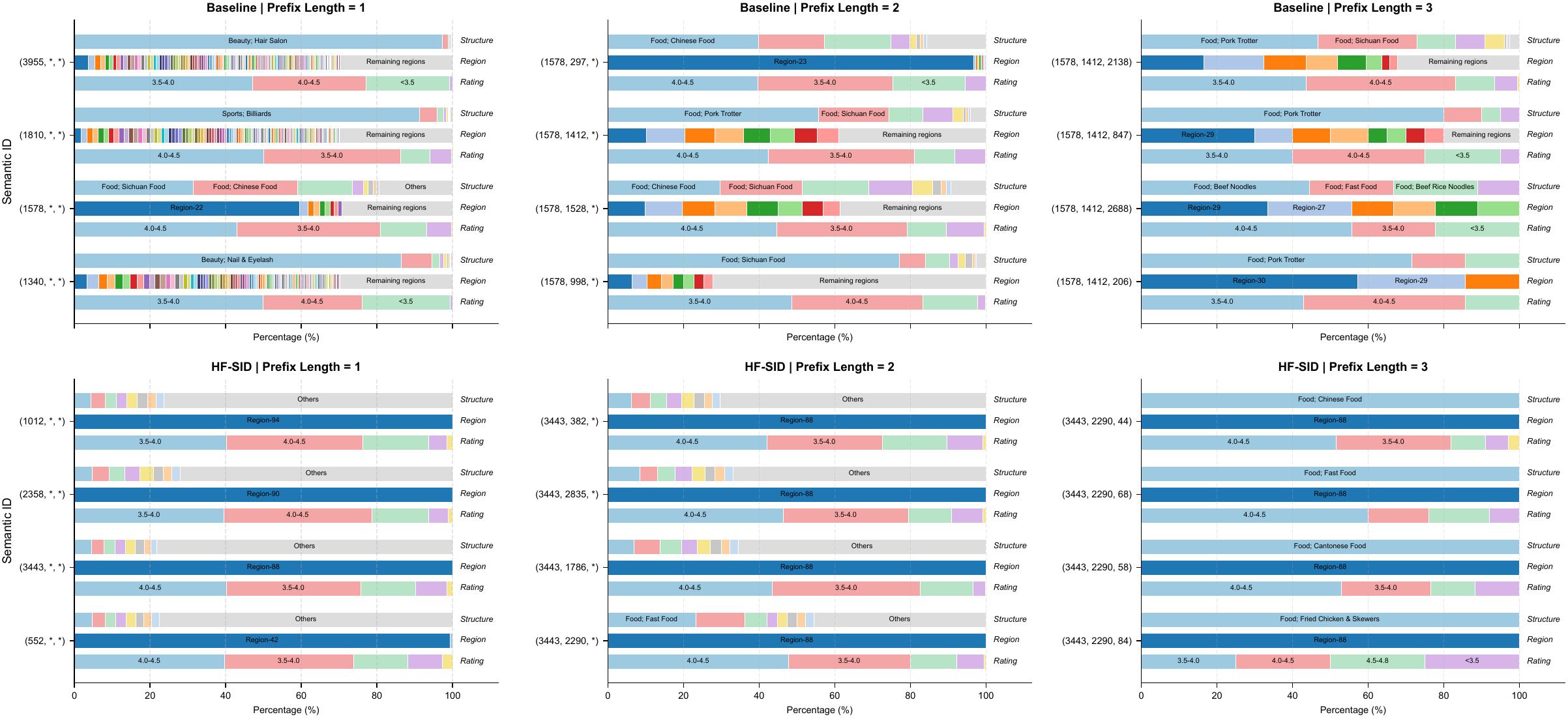}
    \caption{
        Hierarchical semantic consistency of SID prefixes. For prefixes of increasing length $(a,*,*)$, $(a,b,*)$, $(a,b,c)$, we visualize the tag, region, and rating distributions under each. HF-SID shows stronger geographic locality and more concentrated ratings at deeper levels; under a shared parent $(3443,2290,*)$, its L3 children are nearly tag-pure, each a distinct fine-grained category, reflecting Structure-based Contrastive Learning on the third-level residual.
    }   

    \label{fig:sid_hierarchy}
\end{figure*}


\subsubsection{Visualization of Structural Awareness}
Here each POI's structure is its two-level tag of the form ``parent/child'', e.g., ``Food/Chinese Food''. We compare HF-SID with RQ-KMeans by visualizing the tag, region, and rating distributions induced by SID prefixes of increasing length, namely $(a,*)$, $(a,b,*)$, and $(a,b,c)$ (Figure~\ref{fig:sid_hierarchy}). HF-SID shows better geographic and rating organization at all three prefix lengths, while its structural advantage emerges specifically at the third token: under a shared parent prefix $(3443, 2290, *)$, its L3 children are almost tag-pure, each corresponding to a distinct fine-grained food category. This is by design, as Structure-based Contrastive Learning acts only on the third-level residual. The first two levels are left to encode geographic, numerical, and semantic information, geography above all; applying the contrastive objective there would compete with the geographic signal that dominates retrieval, which is precisely the failure the last-layer-only design avoids.

\begin{table}[t]
        \setlength{\abovecaptionskip}{0.2cm}
        \setlength{\belowcaptionskip}{-0.2cm}
    \centering
    \caption{
        Performance comparison on public datasets.
    }
    \label{tab:public_dataset_results}
    \setlength{\tabcolsep}{3.2mm}
    \scalebox{0.80}{
        \begin{tabular}{l l | ccc}
        \hline
        \hline
        \textbf{Dataset}  & \textbf{Method} & \textbf{ICR} ($\uparrow$) & \textbf{Avg.Dist.} ($\downarrow$) & \textbf{Hit@5} ($\uparrow$) \\
        \hline
        \multirow{5}{*}{NYC}
          & RQ-KMeans& 52.13\% & 6.96\,km  & 41.23\% \\
          & GNPR-SID& 19.74\% & 5.52\,km  & 44.34\% \\
          & Cosine-RQ& 16.57\% & 4.61\,km  & 42.87\% \\
          & Pro-GEO& 55.29\% & 6.10\,km  & 53.00\% \\
            & HF-SID& \textbf{56.12\%} & \textbf{3.99\,km} & \textbf{59.83\%} \\
        \hline
        \multirow{5}{*}{TKY}
          & RQ-KMeans& 49.12\% & 7.21\,km  & 30.00\% \\
            & GNPR-SID& 18.50\% & 5.67\,km  & 33.68\% \\
          & Cosine-RQ& 13.69\% & 5.66\,km  & 35.27\% \\
          & Pro-GEO& 45.60\% & 6.19\,km  & 49.27\% \\
        & HF-SID& \textbf{54.46\%} & \textbf{4.41\,km} & \textbf{57.37\%} \\
        \hline
        \hline
        \end{tabular}
    }
    \end{table}

\subsection{Experiments on Public Datasets }
\label{sec:public}
In this section, we evaluate on two public Foursquare check-in datasets, NYC and TKY. Unlike the AMap dataset, these datasets provide only coordinates and check-in records, without the rich dynamic numerical attributes or fine-grained category structure that Num-CPT and Structure-based Contrastive Learning rely on. We therefore apply only Geo-CPT and report results in Table~\ref{tab:public_dataset_results}.

HF-SID achieves the best performance on both datasets across all three metrics. On the SID quality side, Avg.Dist.\ drops from 6.96\,km to 3.99\,km on NYC ($-$42.7\%) and from 7.21\,km to 4.41\,km on TKY ($-$38.8\%), while ICR improves to 56.12\% and 54.46\% respectively---both the highest among all methods. Notably, while GNPR-SID and Cosine-RQ achieve better Avg.Dist.\ than RQ-KMeans on NYC (5.52\,km and 4.61\,km), their ICR collapses to below 20\%, indicating severe codeword collision at the cost of geographic coherence. HF-SID is the only method that improves both metrics simultaneously. On the retrieval side, HF-SID achieves Hit@5 of 59.83\% on NYC and 57.37\% on TKY, outperforming the strongest baseline Pro-GEO by 6.83 and 8.10 points respectively, demonstrating that finer geographic SIDs directly translate to better retrieval accuracy.

The relative Avg.Dist.\ reduction is smaller than on AMap-L ($-$42.7\% vs.\ $-$95.9\%), which is expected: public datasets contain far fewer POIs, so each codebook cluster spans a larger and more geographically dispersed set of POIs, making fine-grained spatial aggregation inherently harder. Even so, consistent gains across two independent cities confirm that Geo-CPT's Cartesian coordinate encoding is robust and platform-agnostic.





\begin{table}[t]
    \setlength{\abovecaptionskip}{0.1cm}
    \setlength{\belowcaptionskip}{-0.2cm}
\centering
\caption{
       Online A/B experiments.}
\label{tab:online_ab_test}
\setlength{\tabcolsep}{1.9mm}
\scalebox{0.80}{
\begin{tabular}{c|ccccc}
    \hline\hline
    Online Metrics & WinRate & PV\_CTR & UV\_CTR & PV\_CVR & UV\_CVR \\
    \hline
    Restaurant         & +0.35\% & +0.24\% & +0.14\% & +5.89\% & +5.63\% \\
    Life Services      & +1.54\% & +0.79\% & +0.23\% & +2.81\% & +2.68\% \\
    Tourist Attraction & +0.87\% & +0.97\% & +0.50\% & +11.51\% & +9.79\% \\
    \hline
    Ave. Imp.
        & \textbf{+0.92\%}& \textbf{+0.67\%}& \textbf{+0.29\%}    & \textbf{+6.74\%}& \textbf{+6.03\%} \\
    \hline \hline
    \end{tabular}
}
\end{table}



\subsection{Online A/B Test }
\label{sec:online}
HF-SID has been deployed on the AMap homepage POI recommendation pipeline, replacing a previously deployed generative retrieval method that already encodes geographic information into its SIDs. The inference pipeline is unchanged: a prompt is assembled from personalized preferences, historical behaviour trajectories, and real-time context including current location, timestamp, and search query; candidates are then generated autoregressively via beam search and passed to the downstream fine-ranking and re-ranking modules. Since the two systems differ only in the SID generation model, the comparison isolates the contribution of HF-SID itself. We ran a one-week A/B test across three representative scenarios, namely Restaurant, Life Services, and Tourist Attraction.

As shown in Table~\ref{tab:online_ab_test}, HF-SID improves every metric in every scenario, and the gains concentrate where fidelity is decisive. Averaged over the three scenarios, PV\_CVR rises by $6.74\%$ and UV\_CVR by $6.03\%$, an order of magnitude beyond the corresponding CTR gains of $0.67\%$ and $0.29\%$. The asymmetry is expected: clicks are driven largely by surface content such as the POI title and cover image, whereas conversion depends on whether the POI is genuinely reachable and genuinely of the intended category, which is exactly what a higher-fidelity identifier preserves. Consistently, Tourist Attraction benefits most with PV\_CVR up $11.51\%$, as it combines the widest geographic span of candidates with dense co-located POIs inside a single scenic area. These gains are obtained over a production system that was already geography-aware.

\section{Conclusion}

In this paper, we present HF-SID, a high-fidelity Semantic ID framework for generative POI retrieval in LBS. HF-SID converts coordinates into a continuous 3D Cartesian form and encodes every numerical value as a single unit through a dedicated Numerical Encoder, fuses these embeddings into the LLM with a Type Embedding that gives each attribute type its own subspace, and aligns the model via Geo-CPT and Num-CPT. A Structure-based Contrastive Learning objective, applied only to the residual before the last quantization layer, then separates co-located POIs that share a coarse tag but differ at the fine level. All mechanisms act at the embedding level and add no token overhead. Deployed on the AMap platform, HF-SID consistently improves both offline SID quality and online business metrics over the previous system, and we release AMap-S, a large-scale real-world POI benchmark. 

\bibliographystyle{ACM-Reference-Format}
\bibliography{WSDM_ref}

\appendix
\section{Appendix}
\subsection{Case Study}
\label{sec:case}

\begin{figure}[t]
    \setlength{\abovecaptionskip}{0.1cm}
    \setlength{\belowcaptionskip}{-0.2cm}
    \centering
    \includegraphics[width=0.98\linewidth]{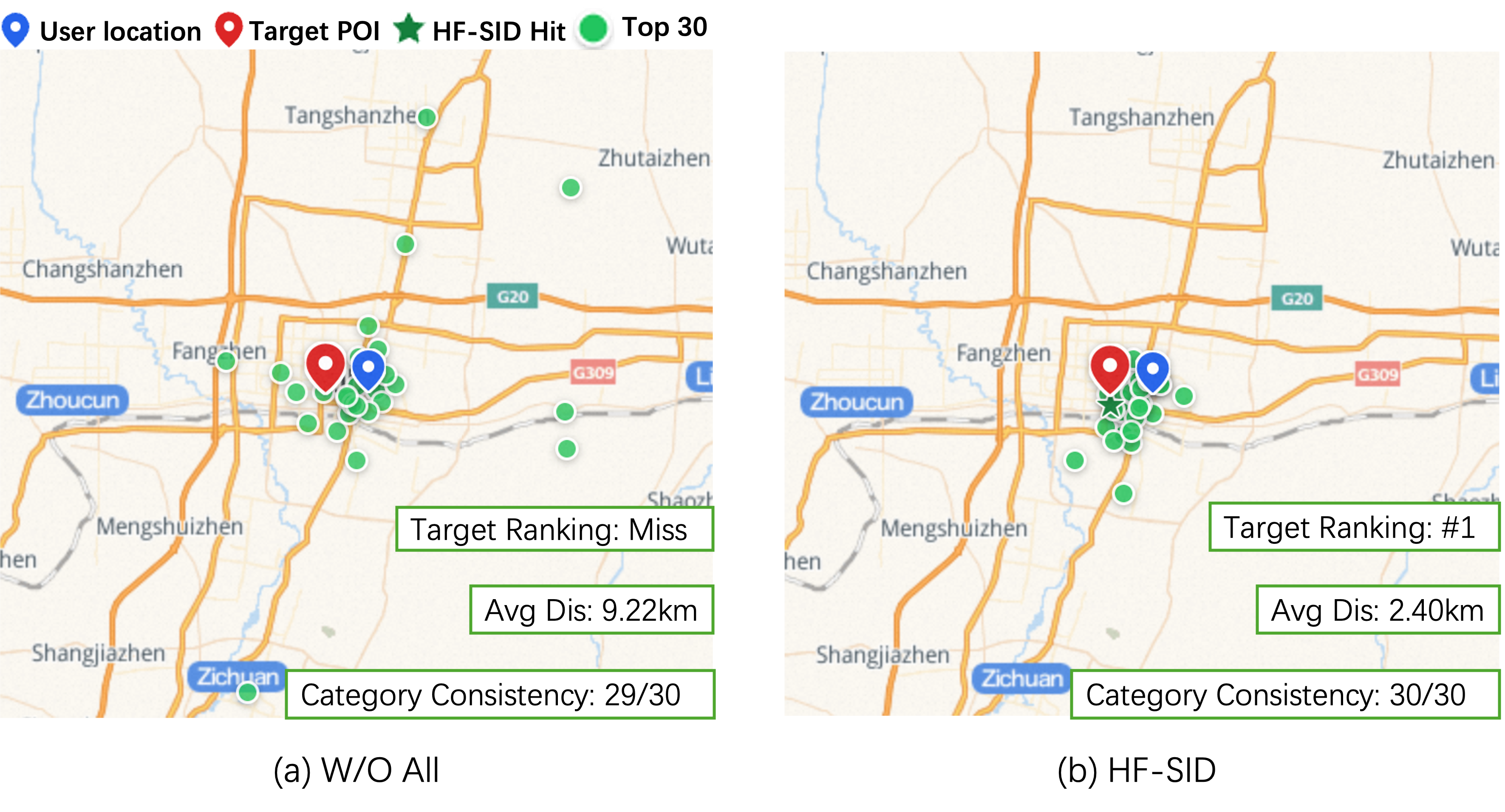}
    \caption{
        Case study of generated POI candidates. The baseline misses the target and produces a geographically dispersed candidate set, whereas HF-SID ranks the target first while achieving lower average distance and higher category consistency among the top-$30$ candidates.
    }
    \label{fig:casestudy}
\end{figure}

We further provide a case study to illustrate how the SID quality improvements translate into concrete retrieval behavior. Figure~\ref{fig:casestudy} compares the generated POI candidates of the baseline and HF-SID under the same user context.

The baseline retrieves candidates that are only partially category-consistent but geographically scattered: although $29$ out of the top-$30$ candidates share the target category, their average distance from the target reaches $9.22$\,km, and the ground-truth target is missed. In contrast, HF-SID concentrates the retrieved candidates around the target region, reduces the average distance to $2.40$\,km, improves category consistency to $30/30$, and ranks the target POI at the first position. This example is consistent with the aggregate results in Table~\ref{tab:main_amap}: preserving geographic and structural fidelity in the representation yields SIDs whose candidate sets are both spatially compact and semantically reliable.

\subsection{CPT Prompt Templates}
\label{app:cpt_templates}

Table~\ref{tab:cpt_templates} summarizes the prompt templates used for continual pre-training. We design two groups of templates corresponding to Geo-CPT and Num-CPT. Geo-CPT focuses on spatial representation and geographic reasoning, including POI spatial information serialization, pairwise distance prediction, and nearest-location identification. Num-CPT focuses on heterogeneous numerical attribute comparison, where the model is required to compare values under a specified attribute type rather than relying on raw numerical magnitude alone.


\begin{table*}[!b]
    \setlength{\abovecaptionskip}{0.05cm}
    \setlength{\belowcaptionskip}{-0.2cm}
    \centering
    \caption{Prompt templates for Geo-CPT and Num-CPT.}
    \label{tab:cpt_templates}

    \setlength{\tabcolsep}{1.0mm}
    \renewcommand{\arraystretch}{0.82}
    \scriptsize

    \scalebox{1}{
    \begin{tabular}{@{}p{0.18\linewidth}|p{0.78\linewidth}@{}}
        \hline
        \hline
        \textbf{Template Type}
        &
        \textbf{Prompt Template}
        \tabularnewline
        \hline

        \raggedright
        \textbf{Prompt 1}\newline
        \textbf{Pairwise Geographic Distance Prediction}
        &
        \raggedright
        The 3D spatial coordinates of \{POI A\} are
        \texttt{<|start\_num\_0|>}\{X\_A\}\texttt{<|end\_num\_0|>},
        \texttt{<|start\_num\_0|>}\{Y\_A\}\texttt{<|end\_num\_0|>}, and
        \texttt{<|start\_num\_0|>}\{Z\_A\}\texttt{<|end\_num\_0|>};
        its address is \{Address A\}.
        The 3D spatial coordinates of \{POI B\} are
        \texttt{<|start\_num\_0|>}\{X\_B\}\texttt{<|end\_num\_0|>},
        \texttt{<|start\_num\_0|>}\{Y\_B\}\texttt{<|end\_num\_0|>}, and
        \texttt{<|start\_num\_0|>}\{Z\_B\}\texttt{<|end\_num\_0|>};
        its address is \{Address B\}.
        The spherical distance between the two locations is
        \texttt{<|start\_num\_1|>}\{Distance\}\texttt{<|end\_num\_1|>}
        meters.
        \tabularnewline
        \hline

        \raggedright
        \textbf{Prompt 2}\newline
        \textbf{Nearest-Location Identification}
        &
        \raggedright
        The 3D spatial coordinates of \{POI A\} are
        \texttt{<|start\_num\_0|>}\{X\_A\}\texttt{<|end\_num\_0|>},
        \texttt{<|start\_num\_0|>}\{Y\_A\}\texttt{<|end\_num\_0|>}, and
        \texttt{<|start\_num\_0|>}\{Z\_A\}\texttt{<|end\_num\_0|>}.
        The 3D spatial coordinates of \{POI B\} are
        \texttt{<|start\_num\_0|>}\{X\_B\}\texttt{<|end\_num\_0|>},
        \texttt{<|start\_num\_0|>}\{Y\_B\}\texttt{<|end\_num\_0|>}, and
        \texttt{<|start\_num\_0|>}\{Z\_B\}\texttt{<|end\_num\_0|>}.
        The 3D spatial coordinates of \{POI C\} are
        \texttt{<|start\_num\_0|>}\{X\_C\}\texttt{<|end\_num\_0|>},
        \texttt{<|start\_num\_0|>}\{Y\_C\}\texttt{<|end\_num\_0|>}, and
        \texttt{<|start\_num\_0|>}\{Z\_C\}\texttt{<|end\_num\_0|>}.
        The location closer to \{Query POI\} is \{Answer POI\}.
        \tabularnewline
        \hline

        \raggedright
        \textbf{Prompt 3}\newline
        \textbf{Attribute Comparison}
        &
        \raggedright
        \{POI A\} has rating
        \texttt{<|start\_num\_2|>}\{Rating\_A\}\texttt{<|end\_num\_2|>},
        weekly views
        \texttt{<|start\_num\_6|>}\{WeekView\_A\}\texttt{<|end\_num\_6|>},
        and collected users
        \texttt{<|start\_num\_5|>}\{Collect\_A\}\texttt{<|end\_num\_5|>}.
        \{POI B\} has rating
        \texttt{<|start\_num\_2|>}\{Rating\_B\}\texttt{<|end\_num\_2|>},
        weekly views
        \texttt{<|start\_num\_6|>}\{WeekView\_B\}\texttt{<|end\_num\_6|>},
        and collected users
        \texttt{<|start\_num\_5|>}\{Collect\_B\}\texttt{<|end\_num\_5|>}.
        \{POI A\} has a higher \{Queried Attribute\} than \{POI B\}.
        \tabularnewline
        \hline
        \hline
    \end{tabular}
    }
\end{table*}

\end{document}